\documentclass[twoside]{articlek}

\usepackage[normal]{caption}

\renewcommand{\refname}{REFERENCES}

\def\BibTeX{{\rm B\kern-.05em{\sc i\kern-.025em b}\kern-.08em
            T\kern-.1667em\lower.7ex\hbox{E}\kern-.125emX}}

\usepackage{amsmath}
\usepackage{graphicx}

\begin{document}

\begin{flushleft}
{\large\bf 
Skyrme RPA for spherical and axially symmetric nuclei}
\vspace*{25pt}

{\bf Anton Repko$^{1,}\footnote{E-mail: \texttt{anton@a-repko.sk}}$, Jan Kvasil$^{1,}\footnote{E-mail: \texttt{kvasil@ipnp.troja.mff.cuni.cz}}$, V.O. Nesterenko$^{2}$ and P.-G. Reinhard$^{3}$}\\
\vspace{5pt}
{$^1$Institute of Particle and Nuclear Physics, Charles University, V Hole\v sovi\v ck\'ach 2, 180 00 Prague 8, Czech Republic\\
$^2$Joint Institute for Nuclear Research, Dubna, Moscow region 141980, Russia\\
$^3$Institut f\"ur Theoretische Physik II, Universit\"at Erlangen, D-91058 Erlangen, Germany}\\

% or
			%{\bf Charlie Somebody$\footnote{E-mail: \texttt{charlie.somebody@gmail.com}}$ and Jack Himself$\footnote{E-mail: \texttt{jack.himself@gmail.com}}$}\\
			%\vspace{5pt}
			%{Institute of Physics, Slovak Academy of Sciences, Dubravska cesta 9, 845 11 Bratislava, Slovakia}\\

% or
			%{\bf Charlie Somebody$^{1}$ and Jack Himself$^{2}$}\\
			%\vspace{5pt}
			%{$^1$Institute of Physics, Slovak Academy of Sciences, Dubravska cesta 9, 845 11 Bratislava, Slovakia\\
			%$^2$FMPI, Comenius University, Mlynska dolina F-1, 842 48 Bratislava, Slovakia}\\

% or
			%{\bf Charlie Somebody and Jack Himself}\\
			%\vspace{5pt}
			%{Institute of Physics, Slovak Academy of Sciences, Dubravska cesta 9, 845 11 Bratislava, Slovakia}\\

\end{flushleft}

%---------------------------------------------------------------------------
\vspace{5pt}
\begin{abstract}\noindent
Random Phase Approximation (RPA) is the basic method for calculation of excited states of nuclei over the Hartree-Fock ground state, suitable also for energy density functionals (EDF or DFT). We developed a convenient formalism for expressing densities and currents in a form of reduced matrix elements, which allows fast calculation of spectra for spherical nuclei. All terms of Skyrme functional were taken into account, so it is possible to calculate electric, magnetic and vortical/toroidal/compression transitions and strength functions of any multipolarity. Time-odd (spin) terms in Skyrme functional become important for magnetic and isovector toroidal (i.e. second-order term in long-wave expansion of E1) transitions. It was also found that transition currents in pygmy region (low-lying part of E1 resonance) exhibit isoscalar toroidal flow, so the previously assumed picture of neutron-skin vibration is not the only mechanism present in pygmy transitions.

RPA calculations with heavy axially-symmetric nuclei now become feasible on ordinary PC. Detailed formulae for axial Skyrme RPA are given. Some numerical results are shown in comparison with the approximate approach of separable RPA, previously developed in our group for fast calculation of strength functions.
\end{abstract}
\vspace{5pt}
%---------------------------------------------------------------------------
\section{INTRODUCTION}
Energy density functional in nuclear physics is a self-consistent microscopic phenomenological approach to calculate nuclear properties and structure over the whole periodic table. The method is analogous to Kohn-Sham density functional theory (DFT) used in electronic systems. Three types of functionals are frequently used nowadays: non-relativistic Skyrme functional \cite{Vautherin1972} with zero-range two-body and density dependent interaction, finite-range Gogny force \cite{Gogny2009} and relativistic (covariant) mean-field \cite{Niksic2014}. Typical approach employs Hartree-Fock-Bogoliubov or HF+BCS calculation scheme to obtain ground state and single-(quasi)particle wavefunctions and energies. These results are then utilized to fit the parameters of the functionals, thus obtaining various parametrizations suitable for specific aims, such as: calculation of mass-table, charge radii, fission barriers, spin-orbit splitting and giant resonances.

Random Phase Approximation (RPA) is a textbook standard \cite{Ring1980} to calculate one-phonon excitations of the nucleus. Increasing computing power has enabled to employ fully self-consistent residual interaction derived from the same density functional as the underlying ground state. While the spherical nuclei can be treated directly (by matrix diagonalization) \cite{Reinhard1992,Terasaki2005,Colo2013}, axially deformed nuclei still pose certain difficulties due to large matrix dimensions \cite{Yoshida2013}. Our group developed a separable RPA (SRPA) approach \cite{Nesterenko2002,Nesterenko2006}, which greatly reduces the computational cost by utilizing separable residual interaction, entirely derived from the underlying functional by means of multi-dimensional linear response theory.

Present article gives a convenient formalism for rotationally-invariant treatment of spherical full RPA, and also gives detailed expressions for matrix elements in axial symmetry. Both time-even and time-odd terms of Skyrme functional are employed, so the method is suitable for various electric and magnetic multipolarities. Finally, some results are shown by means of strength functions and transition currents.

\section{Skyrme functional and density operators}
Skyrme interaction is defined as
\begin{align}
\hat{V}_\mathrm{Sk}(\vec{r}_1,\vec{r}_2) & =
t_0(1+x_0\hat{P}_\sigma)\delta(\vec{r}_1-\vec{r}_2)
-{\textstyle\frac{1}{8}}t_1(1+x_1\hat{P}_\sigma) \big[
(\overleftarrow{\nabla}_1-\overleftarrow{\nabla}_2)^2\delta(\vec{r}_1-\vec{r}_2) + \delta(\vec{r}_1-\vec{r}_2)(\overrightarrow{\nabla}_1-\overrightarrow{\nabla}_2)^2
\big] \nonumber\\
& \quad{}+{\textstyle\frac{1}{4}}t_2(1+x_2\hat{P}_\sigma)(\overleftarrow{\nabla}_1-\overleftarrow{\nabla}_2) \cdot \delta(\vec{r}_1-\vec{r}_2)(\overrightarrow{\nabla}_1-\overrightarrow{\nabla}_2)
+{\textstyle\frac{1}{6}}t_3(1+x_3\hat{P}_\sigma)\delta(\vec{r}_1-\vec{r}_2) \rho^\alpha\big({\textstyle\frac{\vec{r}_1+\vec{r}_2}{2}}\big) \nonumber\\
\label{V_skyrme}
& \quad{}+{\textstyle\frac{\mathrm{i}}{4}}t_4
(\vec{\sigma}_1+\vec{\sigma}_2)\cdot\big[(\overleftarrow{\nabla}_1-\overleftarrow{\nabla}_2) \times \delta(\vec{r}_1-\vec{r}_2)(\overrightarrow{\nabla}_1-\overrightarrow{\nabla}_2)\big]
\end{align}
with parameters $t_0,t_1,t_2,t_3,t_4,x_0,x_1,x_2,x_3,\alpha$ and a spin-exchange operator $\hat{P}_\sigma = \frac{1}{2}(1+\vec{\sigma}_1\cdot\vec{\sigma}_2)$. Since it is a zero-range interaction, the solution of many-body problem by Hartree-Fock method can be equivalently reformulated as a density functional theory \cite{Vautherin1972}, where the complete density functional is
\begin{align}
\mathcal{H}_\mathrm{Sk} & = \frac{1}{2}\sum_{\alpha\beta} \langle \alpha\beta|\hat{V}_\mathrm{Sk}|\alpha\beta\rangle -
\frac{1}{2}\sum_{q=p,n}\sum_{\alpha\beta\in q} \langle \alpha\beta|\hat{V}_\mathrm{Sk}|\beta\alpha\rangle \nonumber\\
& = \int\mathrm{d}^3 r\Big\{ \frac{b_0}{2}\rho^2 - \frac{b_0'}{2}\sum_q\rho_q^2
+b_1(\rho\tau-\vec{j}^2) - b_1'\sum_q(\rho_q\tau_q-\vec{j}_q^2)
+\frac{b_2}{2}(\vec{\nabla}\rho)^2 - \frac{b_2'}{2}\sum_q(\vec{\nabla}\rho_q)^2 \nonumber\\
&\qquad {}+\tilde{b}_1\Big(\vec{s}\cdot\vec{T} - \sum_{ij} \mathcal{J}_{ij}^2\Big)
+\tilde{b}_1'\sum_q\Big(\vec{s}_q\cdot\vec{T}_q-\sum_{ij} \mathcal{J}_{q;ij}^2\Big)
+\frac{b_3}{3}\rho^{\alpha+2}-\frac{b_3'}{3}\rho^\alpha\sum_q\rho_q^2 \nonumber\\
&\qquad {}-b_4\big[\rho\vec{\nabla}\cdot\vec{\mathcal{J}}
+\vec{s}\cdot(\vec{\nabla}\times\vec{j})\big]
-b_4'\sum_q\big[\rho_q\vec{\nabla}\cdot\vec{\mathcal{J}}_q+\vec{s}_q\cdot (\vec{\nabla}\times\vec{j}_q)\big] \nonumber\\
\label{Skyrme_DFT}
&\qquad {}+\frac{\tilde{b}_0}{2}\vec{s}^2-\frac{\tilde{b}_0'}{2}\sum_q\vec{s}_q^2
+\frac{\tilde{b}_2}{2}\sum_{ij}(\nabla_i s_j)^2-\frac{\tilde{b}_2'}{2}\sum_q \sum_{ij}(\nabla_i s_j)_q^2
+\frac{\tilde{b}_3}{3}\rho^\alpha\vec{s}^2-\frac{\tilde{b}_3'}{3}\rho^\alpha\sum_q \vec{s}_q^2 \bigg\}
\end{align}
where the last line contains the spin (time-odd) terms, which are usually omitted. However, they have quite important contribution for magnetic excitations \cite{Vesely2009}, as will be illustrated below, so we use them in all calculations. Parameters $b_j$ depend on the parameters $t_j,x_j$ from (\ref{V_skyrme}):
\begin{equation}
\begin{array}{l}
b_0 = \frac{t_0(2+x_0)}{2}, \qquad b_0' = \frac{t_0(1+2x_0)}{2}, \qquad
\tilde{b}_0 = \frac{t_0 x_0}{2}, \qquad \tilde{b}_0' = \frac{t_0}{2}, \phantom{\Big|}\\
b_1 = \frac{t_1(2+x_1)+t_2(2+x_2)}{8}, \qquad
b_1' = \frac{t_1(1+2x_1)-t_2(1+2x_2)}{8}, \qquad
\tilde{b}_1 = \frac{t_1 x_1 + t_2 x_2}{8}, \qquad
\tilde{b}_1' = \frac{-t_1+t_2}{8}, \phantom{\Big|} \\
b_2 = \frac{3t_1(2+x_1)-t_2(2+x_2)}{16}, \qquad
b_2' = \frac{3t_1(1+2x_1)+t_2(1+2x_2)}{16}, \qquad
\tilde{b}_2 = \frac{3t_1 x_1 - t_2 x_2}{16}, \qquad
\tilde{b}_2' = \frac{3t_1 + t_2}{16}, \phantom{\Big|} \\
b_3 = \frac{t_3(2+x_3)}{8}, \qquad b_3' = \frac{t_3(1+2x_3)}{8}, \qquad
\tilde{b}_3 = \frac{t_3 x_3}{8}, \qquad \tilde{b}_3' = \frac{t_3}{8}, \qquad
b_4 = b_4' = \frac{t_4}{2} \phantom{\Big|}
\end{array}
\end{equation}
Most Skyrme parametrization set explicitly $\tilde{b}_1=\tilde{b}_1'=0$ and this fact is denoted as exclusion of the ``tensor term''. There are also parametrizations fitted with the tensor term included, e.g. SGII \cite{Giai1981}, SLy7 \cite{Chabanat1998}.

The operators corresponding to the densities and currents in (\ref{Skyrme_DFT}) are defined as
\begin{equation}
\label{Jd_op}
\begin{array}{rl}
\textrm{density:}\vphantom{\Big|} & \hat{\rho}(\vec{r}_0) = \delta(\vec{r}-\vec{r}_0), \qquad\qquad
\textrm{kinetic energy:} \quad \hat{\tau}(\vec{r}_0) =
\overleftarrow{\nabla}\cdot\delta(\vec{r}-\vec{r}_0)\overrightarrow{\nabla} \\
\textrm{spin-orbital:}\vphantom{\Big|} & \hat{\mathcal{J}}_{jk}(\vec{r}_0) = \tfrac{\mathrm{i}}{2}
\big[\overleftarrow{\nabla}_j\sigma_k\delta(\vec{r}-\vec{r}_0)
-\delta(\vec{r}-\vec{r}_0)\overrightarrow{\nabla}_j\sigma_k\big] \\
\textrm{vector spin-orbital:}\vphantom{\Big|} & \hat{\vec{\mathcal{J}}}(\vec{r}_0) = \tfrac{\mathrm{i}}{2} \big[\overleftarrow{\nabla}\times\vec{\sigma}\delta(\vec{r}-\vec{r}_0)
-\delta(\vec{r}-\vec{r}_0)\overrightarrow{\nabla}\times\vec{\sigma}\big],\qquad
\hat{\mathcal{J}}_i = \sum_{ijk}\varepsilon_{ijk}\mathcal{J}_{jk} \\
\textrm{current:}\vphantom{\Big|} & \hat{\vec{j}}(\vec{r}_0) = \tfrac{\mathrm{i}}{2}
\big[\overleftarrow{\nabla}\delta(\vec{r}-\vec{r}_0)
-\delta(\vec{r}-\vec{r}_0)\overrightarrow{\nabla}\big],\qquad
\textrm{spin:}\quad \hat{\vec{s}}(\vec{r}_0) = \vec{\sigma}\delta(\vec{r}-\vec{r}_0) \\
\textrm{kinetic energy-spin:}\vphantom{\Big|} & \hat{T}_j(\vec{r}_0) =
\overleftarrow{\nabla}\cdot\sigma_j\delta(\vec{r}-\vec{r}_0)\overrightarrow{\nabla}
\end{array}
\end{equation}
and they are understood as single-particle operators in a many-body system. The densities $\rho,\tau,\mathcal{J}$ are time-even and the currents $\vec{j},\vec{s},\vec{T}$ are time-odd. Spin-orbital current $\mathcal{J}_{ij}$ and current $\nabla_i s_j$ have two indices, so we will decompose them into scalar, vector and (symmetric) tensor part
\begin{equation}
\sum_{ij} \mathcal{J}_{ij}^2 = \frac{1}{3}\sum_i \mathcal{J}_{ii}^2 +
\frac{1}{2}\vec{\mathcal{J}}^2 + \sum_{m=-2}^2 (-1)^m
{[\boldsymbol{\mathcal{J}}_{\!t}^{\vphantom{*}}]}_m {[\boldsymbol{\mathcal{J}}_{\!t}^{\vphantom{*}}]}_{-m} =
\frac{1}{3}\mathcal{J}_s^2 + \frac{1}{2}\vec{\mathcal{J}}^2 + \boldsymbol{\mathcal{J}}_{\!t}^2
\end{equation}

The coefficients related to the (non-relativistic) time-reversal symmetry of the operators are defined as
\begin{equation}
|\bar{\alpha}\rangle = (-1)^{l_\alpha+j_\alpha+m_\alpha}|{-}\alpha\rangle, \quad
u_{\alpha\beta}^{(\pm)} = u_\alpha^{\phantom{|}}v_\beta^{\phantom{|}}\pm v_\alpha^{\phantom{|}} u_\beta^{\phantom{|}}, \quad
\hat{T}^{-1}\hat{A}\hat{T} = \gamma_T^A\hat{A}^\dagger, \quad
\gamma_T^A = \Big\{
\begin{array}{l} +1:\quad\textrm{time-even} \\ -1:\quad\textrm{time-odd}\end{array}
\end{equation}
\begin{equation}
\label{2qp_operator}
\hat{A} = \frac{1}{2}\sum_{\alpha\beta}u_{\alpha\beta}^{(\gamma_T^A)}
\langle\alpha|\hat{A}|\bar{\beta}\rangle
(-\hat{\alpha}_\alpha^+\hat{\alpha}_\beta^+ + \gamma_T^A
\hat{\alpha}_{\bar{\alpha}}^{\phantom{*}}\hat{\alpha}_{\bar{\beta}}^{\phantom{*}})
= \frac{1}{2}\sum_{\alpha\beta}u_{\alpha\beta}^{(\gamma_T^A)}
\langle\bar{\alpha}|\hat{A}|\beta\rangle
(\hat{\alpha}_{\bar{\alpha}}^+\hat{\alpha}_{\bar{\beta}}^+ - \gamma_T^A \hat{\alpha}_\alpha^{\phantom{|}}\hat{\alpha}_\beta^{\phantom{|}})
\end{equation}
where factors $u_\beta,\,v_\beta$ define the Bogoliubov transformation of particles $\hat{a}_\beta^+$ to quasiparticles $\hat{\alpha}_\beta^+$ \cite{Ring1980}. Pairing interaction in the particle-particle channel is defined as ($\rho(\vec{r})$ is the nucleon ground-state density)
\begin{equation}
\label{V_pair}
\hat{V}_\mathrm{pair} = \sum_{q=p,n}\sum_{ij\in q}^{i<j} V_q\delta(\vec{r}_i-\vec{r}_j) \qquad\textrm{or}\qquad
\hat{V}'_\mathrm{pair} = \sum_{q=p,n}\sum_{ij\in q}^{i<j} V_q\Big(1-\frac{\rho(\vec{r}_i)}{\rho_0}\Big)\delta(\vec{r}_i-\vec{r}_j)
\end{equation}
with constant parameters $V_p,\,V_n$ and $\rho_0$. Contribution to the density functional is then
\begin{equation}
\mathcal{H}_\mathrm{pair}
= \langle\mathrm{BCS}|\hat{V}_\mathrm{pair}|\mathrm{BCS}\rangle
= \frac{1}{4}\sum_{q=p,n}V_q\int\kappa_q^2(\vec{r})\mathrm{d}^3 r, \quad
\textrm{optionally with }\times\Big(1-\frac{\rho(\vec{r})}{\rho_0}\Big);
\end{equation}
and the pairing density with the corresponding operator is
\begin{align}
\kappa_q(\vec{r}) &= \sum_{\beta\in q}^{m_\beta>0}
2 f_\beta u_\beta v_\beta \psi_\beta^\dagger(\vec{r})\psi_\beta(\vec{r}) \\
\hat{\kappa}(\vec{r}) &= \sum_{\beta>0} f_\beta\psi_\beta^\dagger(\vec{r})\psi_\beta(\vec{r})
\big[2 u_\beta v_\beta(1-\hat{\alpha}_\beta^+\hat{\alpha}_\beta-\hat{\alpha}_{\bar{\beta}}^+\hat{\alpha}_{\bar{\beta}}) + (u_\beta^2-v_\beta^2)
(\hat{\alpha}_\beta^+\hat{\alpha}_{\bar{\beta}}^+ + \hat{\alpha}_{\bar{\beta}}\hat{\alpha}_\beta) \big]
\end{align}
with energy-cutoff weight $f_\beta$ \cite{Bender2000}. We use monopole pairing, and we involve only diagonal pairs.

\section{RPA formalism in spherical symmetry}
Notation of the Clebsch-Gordan coefficients comes from the book of Varshalovich \cite{Varshalovich1988}. In the language of spherical tensors, a hermitian operator $\hat{A}$ satisfies
\begin{equation}
\label{hermit}
\hat{A}_m^\dagger = (-1)^m\hat{A}_{-m},
\end{equation}
for example a vector (rank-1) operator:
\begin{equation}
\hat{A}_1 = (-\hat{A}_x-\mathrm{i}\hat{A}_y)/\sqrt{2},\quad
\hat{A}_0 = \hat{A}_z,\quad
\hat{A}_{-1} = (\hat{A}_x-\mathrm{i}\hat{A}_y)/\sqrt{2}.
\end{equation}
The position-dependent operators (\ref{Jd_op}), in general denoted as $\hat{\mathbf{J}}_d(\vec{r})$, are as well expressed in terms of quasiparticles (\ref{2qp_operator}) and decomposed in a manner reminiscent of Wigner-Eckart theorem
\begin{equation}
\label{sph_rme}
\hat{\mathbf{J}}_d(\vec{r}) = \frac{1}{2}\sum_{\alpha\beta LJM}
J_{d;\alpha\beta}^{JL}(r)\frac{(-1)^{l_\beta}}{\sqrt{2J+1}}
C_{j_\alpha m_\alpha j_\beta m_\beta}^{JM} \mathbf{Y}_{JM}^{L*}(\vartheta,\varphi)
(-\hat{\alpha}_{\alpha}^+ \hat{\alpha}_{\beta}^+ + \gamma_T^d
\hat{\alpha}_{\bar{\alpha}} \hat{\alpha}_{\bar{\beta}})
\end{equation}
where the symbol $\mathbf{Y}_{JM}^{L*}$ in bold font represents the (complex-conjugated) spherical harmonics in its scalar, vector or tensor form (depending on the rank of operator $\hat{\mathbf{J}}_d(\vec{r})$: $s=0,\,1,\,2$):
\begin{equation}
\label{sph_vectors}
\mathbf{Y}_{JM}^L(\vartheta,\varphi) =
\sum_{m\mu}C_{Lms\mu}^{JM} Y_{Lm}\mathbf{e}_\mu =
\sum_{m} (-1)^m \big[\mathbf{Y}_{JM}^L\big]_m\mathbf{e}_{-m} =
(-1)^{J+L+M+s}\mathbf{Y}_{J,-M}^{L*}(\vartheta,\varphi)
\end{equation}
Expression (\ref{sph_rme}) defines the reduced matrix elements $J_{d;\alpha\beta}^{JL}(r)$, e.g., a standard density matrix element:
\begin{equation}
\rho_{\alpha\beta}^L(r) =
u_{\alpha\beta}^{(+)}\sqrt{(2j_\alpha+1)(2l_\alpha+1)(2j_\beta+1)(2l_\beta+1)}\,
R_\alpha(r) R_\beta(r)
\frac{(-1)^{j_\beta+\frac{1}{2}}}{\sqrt{4\pi}}
\begin{Bmatrix} 
l_\alpha & \!l_\beta\! & L \\ j_\beta & \!j_\alpha\! & \frac{1}{2} \end{Bmatrix}
C_{l_\alpha 0 l_\beta 0}^{L 0}
\end{equation}
and radial wavefunctions are taken from
\begin{equation}
\langle\vec{r}|\alpha\rangle = \psi_\alpha(\vec{r}) =
% R_\alpha(r)\Omega_{j_\alpha m_\alpha}^{l_\alpha}(\vartheta,\varphi) =
R_\alpha(r) \sum_{\nu, s}
C_{l_\alpha,\nu,\frac{1}{2},s}^{j_\alpha,m_\alpha}
Y_{l_\alpha,\nu}^{\phantom{|}}(\vartheta,\varphi)\,
\chi_{s}^{\phantom{|}}
\end{equation}
with spinors $\chi_{\pm1/2} = \binom{1}{0},\,\binom{0}{1}$. Remaining reduced matrix elements will be published later.

Excitations of a given multipolarity are treated as the RPA phonons. One-phonon state, $|\nu\rangle$, is created by action of an operator $\hat{C}_\nu^+$ on the RPA ground state $|\textrm{RPA}\rangle$, and has an excitation energy $E_\nu = \hbar\omega_\nu$.
\begin{equation}
\hat{C}_\nu^+|\textrm{RPA}\rangle=|\nu\rangle,\quad
\hat{C}_\nu^{\phantom{|}}|\textrm{RPA}\rangle=0
\end{equation}
The operator $\hat{C}_\nu^+$ is a two-quasiparticle ($2qp$) operator defined by real coefficients $c_{\alpha\beta}^{(\nu\pm)}$
\begin{equation}
\label{RPA_phon}
\hat{C}_\nu^+ = \frac{1}{2}\sum_{\alpha\beta}
C_{j_\alpha m_\alpha j_\beta m_\beta}^{\lambda_\nu\mu_\nu}\Big(
c_{\alpha\beta}^{(\nu-)}\hat{\alpha}_{\alpha}^+\hat{\alpha}_{\beta}^+ +
c_{\alpha\beta}^{(\nu+)}\hat{\alpha}_{\bar{\alpha}}^{\phantom{*}}
\hat{\alpha}_{\bar{\beta}}^{\phantom{*}} \Big)
\end{equation}
and satisfies the RPA equation
\begin{equation}
\label{RPA_eq}
{[\hat{H},\hat{C}_\nu^+]}_{2qp} = E_\nu\hat{C}_\nu^+
\end{equation}
where the index $2qp$ means that we take only the two-quasiparticle portion of the commutator (after normal ordering). Although all commutators should be evaluated in the RPA ground state, it is common to evaluate them in the HF+BCS ground state (i.e., we use quasi-boson approximation), since the contribution of $4qp$ and higher correlations in the ground state is assumed to be low. 

Formulae (\ref{sph_rme}) and (\ref{RPA_phon}) involve duplicate $2qp$ pairs. To remove them consistently, we will rescale diagonal pairing factors
\begin{equation}
\label{order2qp}
\frac{1}{2}\sum_{\alpha\beta}\mapsto\sum_{\alpha\geq\beta},\quad
u_{\alpha\alpha}^{(+)} = \sqrt{2}\,u_\alpha v_\alpha\quad\textrm{(instead of $2u_\alpha v_\alpha$)}
\end{equation}
and $c_{\alpha\alpha}^{(\nu\pm)}$ will be rescaled automatically. Diagonal matrix elements contribute only to electric transitions with $\lambda$ even.

The hamiltonian is taken as a sum of mean-field part (HF+BCS) and the second functional derivative of Skyrme density functional (+ Coulomb and pairing interaction).
\begin{equation}
\hat{H} = \hat{H}_0 + \hat{V}_\mathrm{res} =
\sum_\gamma\varepsilon_\gamma\hat{\alpha}_\gamma^+\hat{\alpha}_\gamma^{\phantom{|}}
+ \frac{1}{2}\sum_{dd'}\int\!\!\!\int\mathrm{d}^3 r_1\,\mathrm{d}^3 r_2
\frac{\delta^2\mathcal{H}}{\delta J_d(\vec{r}_1)\delta J_{d'}(\vec{r}_2)}
:\!\hat{J}_{d}(\vec{r}_1)\hat{J}_{d'}(\vec{r}_2)\!:
\end{equation}
RPA equation (\ref{RPA_eq}) then turns into a matrix equation
\begin{equation}
\label{fullRPA_eq}
\begin{pmatrix} A & B \\ B & A \end{pmatrix} \binom{c^{(\nu-)}}{c^{(\nu+)}} =
\begin{pmatrix} E_\nu & 0 \\ 0 & -E_\nu \end{pmatrix}
\binom{c^{(\nu-)}}{c^{(\nu+)}}
\end{equation}
which can be recast into half-dimensional symmetric-matrix eigenvalue problem \cite{Ring1980}. Matrices $A$ and $B$ are
\begin{subequations}
\begin{align}
\label{fullRPA_A}
A_{pp'} & = \delta_{pp'}\varepsilon_p + \sum_{dd'L}\frac{(-1)^{l_\beta+l_\delta}}{2\lambda+1}
\int_0^\infty \frac{\delta^2\mathcal{H}}{\delta J_d\delta J_{d'}}
J_{d;p}^{\lambda L}(r)J_{d';p'}^{\lambda L*}(r)r^2\mathrm{d}r \\
\label{fullRPA_B}
B_{pp'} & = \sum_{dd'L} \gamma_T^d \frac{(-1)^{l_\beta+l_\delta}}{2\lambda+1}
\int_0^\infty \frac{\delta^2\mathcal{H}}{\delta J_d\delta J_{d'}}
J_{d;p}^{\lambda L}(r)J_{d';p'}^{\lambda L*}(r)
r^2\mathrm{d}r,\qquad (p\equiv\alpha\beta,\,p'\equiv\gamma\delta)
\end{align}
\end{subequations}
The expression $\frac{\delta^2\mathcal{H}}{\delta J_d\delta J_{d'}}$ is symbolical, and includes the integration of the delta function, yielding $\vec{r}_1 = \vec{r}_2$. The exchange Coulomb interaction can be treated by Slater approximation as a density functional
\begin{equation}
\mathcal{H}_\mathrm{xc} = -\frac{3}{4}\bigg(\frac{3}{\pi}\bigg)^{\!1/3}
\frac{e^2}{4\pi\epsilon_0}\int\mathrm{d}^3 r \rho_p^{4/3}(\vec{r})
\end{equation}
however, the direct Coulomb interaction gives rise to a double integral instead
\begin{align}
\sum_L\int_0^\infty\frac{\delta^2\mathcal{H}}{\delta J_d\delta J_{d'}}
J_{d;\alpha\beta}^{\lambda L*}(r)J_{d';\gamma\delta}^{\lambda L}(r)
r^2\mathrm{d}r \quad\mapsto \nonumber\\
\frac{e^2}{4\pi\epsilon_0} \frac{4\pi}{2\lambda+1}
\int_0^\infty r^2\mathrm{d}r &\int_0^\infty r'^2\mathrm{d}r' 
\rho_{\alpha\beta}^\lambda(r)\rho_{\gamma\delta}^\lambda(r')
\times \bigg\{\!\!\begin{array}{l}
r^\lambda / r'^{\lambda+1} \quad(r<r') \phantom{\big|}\\
r'^\lambda / r^{\lambda+1} \quad(r\geq r')\phantom{\big|} \end{array}
\end{align}

After calculation of the RPA states, yielding $E_\nu$ and $c_{\alpha\beta}^{(\nu\pm)}$, we are interested in the matrix elements of electric and magnetic transition operators and in the transition densities and currents.
\begin{align}
\label{trans_me}
\langle\nu|\hat{M}_{\lambda\mu}|\textrm{RPA}\rangle & =
\langle[\hat{C}_\nu^{\phantom{|}},\hat{M}_{\lambda\mu}]\rangle =
\sum_{\alpha\geq\beta} \frac{(-1)^{l_\beta+1}}{\sqrt{2\lambda+1}}
M_{\lambda;\alpha\beta}
\Big( c_{\alpha\beta}^{(\nu-)} + \gamma_T^M c_{\alpha\beta}^{(\nu+)} \Big)^{\!*} \\
\delta\rho_{q;\nu}(\vec{r}) & =
\langle[\hat{C}_\nu^{\phantom{|}},\hat{\rho}_{q}(\vec{r})]\rangle =
\sum_{\alpha\geq\beta} \frac{(-1)^{l_\beta+1}}{\sqrt{2\lambda+1}}
\rho_{q;\alpha\beta}^\lambda(r)
\Big( c_{\alpha\beta}^{(\nu-)} + c_{\alpha\beta}^{(\nu+)} \Big)^{\!*}
Y_{\lambda\mu}^*(\vartheta,\varphi) \\
\delta\vec{j}_{q;\nu}(\vec{r}) & =
\langle[\hat{C}_\nu^{\phantom{|}},\hat{\vec{j}}_{q}(\vec{r})]\rangle =
\sum_L\sum_{\alpha\geq\beta} \frac{(-1)^{l_\beta+1}}{\sqrt{2\lambda+1}}
j_{q;\alpha\beta}^{\lambda L}(r)
\Big( c_{\alpha\beta}^{(\nu-)} - c_{\alpha\beta}^{(\nu+)} \Big)^{\!*}
\vec{Y}_{\lambda\mu}^{L*}(\vartheta,\varphi)
\end{align}
Besides electric ($\gamma_T^{\mathrm{E}\lambda} = 1$) and magnetic ($\gamma_T^{\mathrm{M}\lambda} = -1$) operators in long-wave approximation, we implement also electric vortical, toroidal and compression operators \cite{Kvasil2011} (toroidal operator constitutes the next-order term in long-wave expansion of the exact electric transition operator)
\begin{subequations}
\label{tran}
\begin{align}
\label{M_E}
\hat{M}_{\lambda\nu}^\mathrm{E} & = \sum_i \hat{M}^{\mathrm{E}}_{\lambda \mu}(\vec{r}_i)
= e\sum_{q=p,n} z_q\sum_{i\in q}
\Big( r^\lambda Y_{\lambda\mu}(\vartheta,\varphi) \Big)_i \\
\hat{M}_{\lambda\nu}^\mathrm{M} & = \frac{\mu_N}{c}\sqrt{\lambda(2\lambda+1)}\,
\sum_{q=p,n} \sum_{i\in q} \bigg(
\bigg[ \frac{g_{s,q}}{2}\vec{\sigma}
+ \frac{2 g_{l,q}}{\lambda+1} \hat{\vec{L}}\,\bigg]\,
r^{\lambda-1}\vec{Y}_{\lambda\mu}^{\lambda-1}(\vartheta,\varphi)
\bigg)_i
\end{align}
\begin{align}
\hat{M}_{\mathrm{vor};\lambda\mu}^\mathrm{E} & = \frac{-\mathrm{i}/c}{2\lambda+3}
\sqrt{\frac{2\lambda+1}{\lambda+1}}\int\mathrm{d}^3 r\,
\hat{\vec{j}}_\mathrm{nuc}(\vec{r})r^{\lambda+1}
\vec{Y}_{\lambda\mu}^{\lambda+1}(\vartheta,\varphi) =
\hat{M}_{\mathrm{tor};\lambda\mu}^\mathrm{E} + \hat{M}_{\mathrm{com};\lambda\mu}^E \\
\label{M_tor}
\hat{M}_{\mathrm{tor};\lambda\mu}^\mathrm{E} & =
\frac{-1}{2c(2\lambda+3)}\sqrt{\frac{\lambda}{\lambda+1}}
\int\mathrm{d}^3 r\,\hat{\vec{j}}_\mathrm{nuc}(\vec{r})\cdot\vec{\nabla}\times
\big[r^{\lambda+2}\vec{Y}_{\lambda\mu}^\lambda(\vartheta,\varphi)\big] \\
% \frac{-\mathrm{i}\sqrt{\lambda}}{2\sqrt{2\lambda+1}} \int\mathrm{d}^3 r\,
% \hat{\vec{j}}_\mathrm{nuc}(\vec{r})\cdot r^{\lambda+1} \bigg[
% \vec{Y}_{\lambda\mu}^{\lambda-1}(\vartheta,\varphi) +
% \sqrt{\frac{\lambda}{\lambda+1}}\frac{2
% \vec{Y}_{\lambda\mu}^{\lambda+1}(\vartheta,\varphi)}{2\lambda+3}\bigg] \\
\label{M_com}
\hat{M}_{\mathrm{com};\lambda\mu}^\mathrm{E} & =
\frac{\mathrm{i}}{2c(2\lambda+3)}
\int\mathrm{d}^3 r\,\hat{\vec{j}}_\mathrm{nuc}(\vec{r})\cdot
\vec{\nabla}\big[r^{\lambda+2} Y_{\lambda\mu}(\vartheta,\varphi)\big]\qquad
\big(= -k\hat{M}_{\mathrm{com};\lambda\mu}^{\mathrm{E}\,\prime}\big) \\
% \frac{\mathrm{i}\sqrt{\lambda}}{2\sqrt{2\lambda+1}} \int\mathrm{d}^3 r\,
% \hat{\vec{j}}_\mathrm{nuc}(\vec{r})\cdot r^{\lambda+1} \bigg[
% \vec{Y}_{\lambda\mu}^{\lambda-1}(\vartheta,\varphi) -
% \sqrt{\frac{\lambda+1}{\lambda}}\frac{2
% \vec{Y}_{\lambda\mu}^{\lambda+1}(\vartheta,\varphi)}{2\lambda+3}\bigg]
\hat{M}_{\mathrm{com};\lambda\mu}^{\mathrm{E}\,\prime} & =
\sum_i \hat{M}_{\mathrm{com};\lambda\mu}^{\mathrm{E}\,\prime}(\vec{r}_i) =
\frac{e}{2(2\lambda+3)} 
\sum_{q=p,n} z_q \sum_{i\in q} \Big( r^{\lambda+2} Y_{\lambda\mu}(\vartheta,\varphi) \Big)_i
\end{align}
\end{subequations}
where $z_q$ are effective charges of the nucleons, $g_{l/s,q}$ are orbital/spin g-factors (we take $g_{l,q} = z_q$; spin g-factors are reduced by a quenching factor $\varsigma=0.7$) and $\vec{j}_\mathrm{nuc}$ is the nuclear current composed of convective and magnetization part
\begin{equation}
\hat{\vec{j}}_\mathrm{nuc}(\vec{r}) = \frac{e\hbar}{m_p}\sum_{q=p,n}\sum_{i\in q}
\Big[ z_q\hat{\vec{j}}_i(\vec{r}) + \frac{1}{4} g_{s,q}\vec{\nabla}\times\hat{\vec{s}}_i(\vec{r})\Big]
\end{equation}
where (convective) current and spin one-body operators are the same as in Skyrme functional (\ref{Jd_op}).

\section{RPA formalism in axial symmetry}
Axial coordinates are
\begin{equation}
\varrho=\sqrt{x^2+y^2},\,z,\,\varphi;\qquad x=\varrho\cos\varphi,\ y=\varrho\sin\varphi
\end{equation}
Calculations in axially deformed nuclei don't conserve total angular momentum (previously denoted $J$ or $\lambda$), nevertheless, they conserve its $z$-projection ($\mu$) and parity, so it is convenient to preserve part of the formalism from spherical symmetry, namely the convention of $m$-components for vector and tensor operators, and their hermitian conjugation (\ref{hermit}). Operators of differentiation are then
\begin{equation}
\nabla_{\pm1} = \frac{1}{\sqrt{2}}\bigg({\mp}\frac{\partial}{\partial x}-\mathrm{i}\frac{\partial}{\partial y}\bigg)
= {\mp}\frac{\mathrm{e}^{\pm\mathrm{i}\varphi}}{\sqrt{2}}
\bigg( \frac{\partial}{\partial\varrho}
 \pm \frac{\mathrm{i}}{\varrho}\frac{\partial}{\partial\varphi} \bigg),\qquad
\nabla_0 = \frac{\partial}{\partial z}
\end{equation}
Single-particle wavefunction (and its time-reversal conjugate) is expressed as a spinor
\begin{equation}
\psi_\alpha(\vec{r}) =
\binom{R_{\alpha\uparrow}(\varrho,z)\mathrm{e}^{\mathrm{i}m_\alpha^-\varphi}}
{R_{\alpha\downarrow}(\varrho,z)\mathrm{e}^{\mathrm{i}m_\alpha^+\varphi}}, \quad
\psi_{\bar{\alpha}}(\vec{r}) = \binom{R_{\alpha\downarrow}(\varrho,z)\mathrm{e}^{-\mathrm{i}m_\alpha^+\varphi}}
{-R_{\alpha\uparrow}(\varrho,z)\mathrm{e}^{-\mathrm{i}m_\alpha^-\varphi}},\quad
\textrm{where }\ m_\alpha^\pm = m_\alpha \pm \frac{1}{2}
\end{equation}
and the radial parts of its derivatives will be denoted by a shorthand notation
\begin{equation}
\begin{split}
\nabla_{\pm1}\psi_\alpha &= {\mp}\frac{\mathrm{e}^{\pm\mathrm{i}\varphi}}{\sqrt{2}}
\binom{(\partial_\varrho R_{\alpha\uparrow} \mp m_\alpha^- R_{\alpha\uparrow}/\varrho)\mathrm{e}^{\mathrm{i}m_\alpha^-\varphi}}
{(\partial_\varrho R_{\alpha\downarrow} \mp m_\alpha^+ R_{\alpha\downarrow}/\varrho)\mathrm{e}^{\mathrm{i}m_\alpha^+\varphi}}
\equiv \mathrm{e}^{\pm\mathrm{i}\varphi}
\begin{pmatrix} R_{\alpha\uparrow}^{(\pm)}\mathrm{e}^{\mathrm{i}m_\alpha^-\varphi} \\
R_{\alpha\downarrow}^{(\pm)}\mathrm{e}^{\mathrm{i}m_\alpha^+\varphi}
\end{pmatrix} \\
\nabla_0\psi_\alpha &=
\binom{\partial_z R_{\alpha\uparrow}\mathrm{e}^{\mathrm{i}m_\alpha^-\varphi}}
{\partial_z R_{\alpha\downarrow}\mathrm{e}^{\mathrm{i}m_\alpha^+\varphi}}
\equiv \begin{pmatrix} R_{\alpha\uparrow}^{(0)}\mathrm{e}^{\mathrm{i}m_\alpha^-\varphi} \\
R_{\alpha\downarrow}^{(0)}\mathrm{e}^{\mathrm{i}m_\alpha^+\varphi} \end{pmatrix}
\end{split}
\end{equation}

Radial functions $R_{\alpha\uparrow\!\downarrow}(\varrho,z),\,R_{\alpha\uparrow\!\downarrow}^{(\pm)}(\varrho,z)$ are real, and their spinor-wise products will be denoted by a dot to keep the expressions simple:
\begin{equation}
R_\alpha \cdot R_\beta \ \equiv\ R_{\alpha\uparrow}(\varrho,z) R_{\beta\uparrow}(\varrho,z)
+ R_{\alpha\downarrow}(\varrho,z) R_{\beta\downarrow}(\varrho,z)
\end{equation}
Vector currents will be decomposed in the style of rank-1 tensor operators. Vector product in the expression for spin-orbital current leads to (for vector product in $m$-scheme see \cite[(1.2.28)]{Varshalovich1988})
\begin{equation}
(\vec{\nabla}\times\vec{\sigma})\psi_\alpha = \left\{\begin{array}{rl}
{+}1: & \mathrm{i}\,\mathrm{e}^{\mathrm{i}\varphi}
\begin{pmatrix} \big({-}R_{\alpha\uparrow}^{(+)} -\sqrt{2}\,R_{\alpha\downarrow}^{(0)}\big)
\mathrm{e}^{\mathrm{i}m_\alpha^-\varphi} \\
R_{\alpha\downarrow}^{(+)}\mathrm{e}^{\mathrm{i}m_\alpha^+\varphi} \end{pmatrix} \\
0: & \mathrm{i}\,\begin{pmatrix}
{-}\sqrt{2}\,R_{\alpha\downarrow}^{(-)}\mathrm{e}^{\mathrm{i}m_\alpha^-\varphi} \\
{-}\sqrt{2}\,R_{\alpha\uparrow}^{(+)}\mathrm{e}^{\mathrm{i}m_\alpha^+\varphi}
\end{pmatrix} \\
{-}1: & \mathrm{i}\,\mathrm{e}^{-\mathrm{i}\varphi}
\begin{pmatrix} R_{\alpha\uparrow}^{(-)}\mathrm{e}^{\mathrm{i}m_\alpha^-\varphi} \\
\big({-}R_{\alpha\downarrow}^{(-)} -\sqrt{2}\,R_{\alpha\uparrow}^{(0)}\big)
\mathrm{e}^{\mathrm{i}m_\alpha^+\varphi} \end{pmatrix} \end{array} \right.
\end{equation}

Matrix elements of densities and currents are then
\begin{subequations}
\label{axial_me}
\begin{align}
\langle\alpha|\hat{\rho}|\beta\rangle &= R_\alpha \cdot R_\beta
\,\mathrm{e}^{\mathrm{i}(m_\beta-m_\alpha)\varphi} \\
\langle\alpha|\hat{\tau}|\beta\rangle &= \big( R_\alpha^{(0)} \cdot R_\beta^{(0)}
+ R_\alpha^{(+)} \cdot R_\beta^{(+)} + R_\alpha^{(-)} \cdot R_\beta^{(-)} \big)
\,\mathrm{e}^{\mathrm{i}(m_\beta-m_\alpha)\varphi}
\end{align}
Factor $\mathrm{e}^{\mathrm{i}(m_\beta-m_\alpha)\varphi}$ will be omitted in the following expressions.
\begin{small}
\begin{align}
\langle\alpha|\vec{\mathcal{J}}|\beta\rangle &= \Bigg\{\!\! \begin{array}{rl}
{+}1:\!& \tfrac{1}{2}\mathrm{e}^{\mathrm{i}\varphi}\big[
\big(R_{\alpha\downarrow}^{(-)} + \sqrt{2}\,R_{\alpha\uparrow}^{(0)}\big)R_{\beta\downarrow}
- R_{\alpha\uparrow}^{(-)} R_{\beta\uparrow}
- R_{\alpha\uparrow}\big(R_{\beta\uparrow}^{(+)}+\sqrt{2}\,R_{\beta\downarrow}^{(0)}\big)
+ R_{\alpha\downarrow} R_{\beta\downarrow}^{(+)} \big] \\
0:\!& \tfrac{1}{2}\big[{-}\sqrt{2}\,\big( R_{\alpha\downarrow}^{(-)} R_{\beta\uparrow} + R_{\alpha\uparrow}^{(+)}R_{\beta\downarrow}
+ R_{\alpha\uparrow}R_{\beta\downarrow}^{(-)} + R_{\alpha\downarrow}R_{\beta\uparrow}^{(+)} \big) \big] \\
{-}1:\!& \tfrac{1}{2}\mathrm{e}^{-\mathrm{i}\varphi}\big[
\big( R_{\alpha\uparrow}^{(+)} + \sqrt{2}\,R_{\alpha\downarrow}^{(0)} \big) R_{\beta\uparrow}
- R_{\alpha\downarrow}^{(+)} R_{\beta\downarrow}
- R_{\alpha\downarrow} \big( R_{\beta\downarrow}^{(-)}+\sqrt{2}\,R_{\beta\uparrow}^{(0)} \big)
+ R_{\alpha\uparrow} R_{\beta\uparrow}^{(-)} \big]
\end{array} \\
\langle\alpha|\vec{j}|\beta\rangle &= \Bigg\{\!\! \begin{array}{rl}
{+}1:\!& \tfrac{\mathrm{i}}{2}\mathrm{e}^{\mathrm{i}\varphi}
\big( {-}R_\alpha^{(-)}\cdot R_\beta - R_\alpha\cdot R_\beta^{(+)} \big) \\
0:\!& \tfrac{\mathrm{i}}{2}
\big( R_\alpha^{(0)}\cdot R_\beta - R_\alpha\cdot R_\beta^{(0)} \big) \\
{-}1:\!& \tfrac{\mathrm{i}}{2}\mathrm{e}^{-\mathrm{i}\varphi}
\big( {-}R_\alpha^{(+)}\cdot R_\beta - R_\alpha\cdot R_\beta^{(-)} \big) \end{array}
\qquad \langle\alpha|\vec{s}|\beta\rangle = \Bigg\{\!\! \begin{array}{rl}
{+}1:\!& \mathrm{e}^{\mathrm{i}\varphi}\big({-}\sqrt{2}\,R_{\alpha\uparrow} R_{\beta\downarrow}\big) \\
0:\!& R_{\alpha\uparrow} R_{\beta\uparrow} - R_{\alpha\downarrow} R_{\beta\downarrow} \\
{-}1:\!& \mathrm{e}^{-\mathrm{i}\varphi}\big(\sqrt{2}\,R_{\alpha\downarrow} R_{\beta\uparrow}\big)
\end{array} \\
\langle\alpha|\vec{T}|\beta\rangle &= \Bigg\{\!\! \begin{array}{rl}
{+}1:\!& \mathrm{e}^{\mathrm{i}\varphi}(-\sqrt{2}\,)
\big[ R_{\alpha\uparrow}^{(0)} R_{\beta\downarrow}^{(0)}
+ R_{\alpha\uparrow}^{(+)} R_{\beta\downarrow}^{(+)}
+ R_{\alpha\uparrow}^{(-)} R_{\beta\downarrow}^{(-)} \big] \\
0:\!& R_{\alpha\uparrow}^{(0)} R_{\beta\uparrow}^{(0)} - R_{\alpha\downarrow}^{(0)} R_{\beta\downarrow}^{(0)}
+ R_{\alpha\uparrow}^{(+)} R_{\beta\uparrow}^{(+)} - R_{\alpha\downarrow}^{(+)} R_{\beta\downarrow}^{(+)}
+ R_{\alpha\uparrow}^{(-)} R_{\beta\uparrow}^{(-)} - R_{\alpha\downarrow}^{(-)} R_{\beta\downarrow}^{(-)} \\
{-}1:\!& \mathrm{e}^{-\mathrm{i}\varphi}\sqrt{2}\,
\big[ R_{\alpha\downarrow}^{(0)} R_{\beta\uparrow}^{(0)}
+ R_{\alpha\downarrow}^{(+)} R_{\beta\uparrow}^{(+)}
+ R_{\alpha\downarrow}^{(-)} R_{\beta\uparrow}^{(-)} \big] \end{array} \\
\langle\alpha|\vec{\nabla}\times\vec{j}|\beta\rangle
&= -\mathrm{i}\big(\vec{\nabla}\psi_\alpha\big)^\dagger\times\vec{\nabla}\psi_\beta
= \Bigg\{\!\! \begin{array}{rl}
{+}1:\!& \mathrm{e}^{\mathrm{i}\varphi}\big( R_\alpha^{(-)} \cdot R_\beta^{(0)}
+ R_\alpha^{(0)} \cdot R_\beta^{(+)} \big) \\
0:\!& R_\alpha^{(-)} \cdot R_\beta^{(-)} - R_\alpha^{(+)} \cdot R_\beta^{(+)} \\
{-}1:\!& \mathrm{e}^{-\mathrm{i}\varphi}\big( {-}R_\alpha^{(+)} \cdot R_\beta^{(0)}
- R_\alpha^{(0)} \cdot R_\beta^{(-)} \big) \end{array} \\
\langle\alpha|\vec{\nabla}\cdot\vec{\mathcal{J}}|\beta\rangle
&= {-}R_{\alpha\uparrow}^{(+)} R_{\beta\uparrow}^{(+)} + R_{\alpha\downarrow}^{(+)} R_{\beta\downarrow}^{(+)}
+ R_{\alpha\uparrow}^{(-)} R_{\beta\uparrow}^{(-)} - R_{\alpha\downarrow}^{(-)} R_{\beta\downarrow}^{(-)} \nonumber\\
& \quad{}-\sqrt{2}\,\big( R_{\alpha\uparrow}^{(0)} R_{\beta\downarrow}^{(-)} + R_{\alpha\downarrow}^{(-)} R_{\beta\uparrow}^{(0)}
+ R_{\alpha\downarrow}^{(0)} R_{\beta\uparrow}^{(+)} + R_{\alpha\uparrow}^{(+)} R_{\beta\downarrow}^{(0)} \big) \\
\langle\alpha|\mathcal{J}_s|\beta\rangle &= \tfrac{\mathrm{i}}{2}\big[
\big( R_{\alpha\uparrow}^{(0)} + \sqrt{2}\,R_{\alpha\downarrow}^{(-)} \big) R_{\beta\uparrow}
- \big( R_{\alpha\downarrow}^{(0)} + \sqrt{2}\,R_{\alpha\uparrow}^{(+)} \big) R_{\beta\downarrow} \nonumber\\
&\qquad {}- R_{\alpha\uparrow} \big( R_{\beta\uparrow}^{(0)} + \sqrt{2}\,R_{\beta\downarrow}^{(-)} \big)
+ R_{\alpha\downarrow} \big( R_{\beta\downarrow}^{(0)} + \sqrt{2}\,R_{\beta\uparrow}^{(+)} \big)\big] \\[5pt]
\langle\alpha|\mathcal{J}_t|\beta\rangle &= \left\{\!\! \begin{array}{rl}
{+}2:\!& \tfrac{\mathrm{i}}{\sqrt{2}} \mathrm{e}^{2\mathrm{i}\varphi}
\big( R_{\alpha\uparrow}^{(-)} R_{\beta\downarrow}
+ R_{\alpha\uparrow} R_{\beta\downarrow}^{(+)} \big) \\
{+}1:\!& \tfrac{\mathrm{i}}{2\sqrt{2}} \mathrm{e}^{\mathrm{i}\varphi}
\big[ {-}R_{\alpha\uparrow}^{(-)} R_{\beta\uparrow}
+ \big(R_{\alpha\downarrow}^{(-)}
 -\sqrt{2}\,R_{\alpha\uparrow}^{(0)} \big)R_{\beta\downarrow}
- R_{\alpha\uparrow}\big(R_{\beta\uparrow}^{(+)}
 -\sqrt{2}\,R_{\beta\downarrow}^{(0)}\big)
+ R_{\alpha\downarrow}R_{\beta\downarrow}^{(+)} \big] \\
0:\!& \tfrac{\mathrm{i}}{2\sqrt{3}}\big[
\big(\sqrt{2}\,R_{\alpha\uparrow}^{(0)}-R_{\alpha\downarrow}^{(-)}\big)R_{\beta\uparrow}
-\big(\sqrt{2}\,R_{\alpha\downarrow}^{(0)}-R_{\alpha\uparrow}^{(+)}\big)R_{\beta\downarrow} \\
 & \qquad{}-R_{\alpha\uparrow}\big(\sqrt{2}\,R_{\beta\uparrow}^{(0)}-R_{\beta\downarrow}^{(-)}\big)
+R_{\alpha\downarrow}\big(\sqrt{2}\,R_{\beta\downarrow}^{(0)}-R_{\beta\uparrow}^{(+)}\big)
\big] \\
{-}1:\!& \tfrac{\mathrm{i}}{2\sqrt{2}} \mathrm{e}^{-\mathrm{i}\varphi}
\big[ {-}\big(R_{\alpha\uparrow}^{(+)}
 -\sqrt{2}\,R_{\alpha\downarrow}^{(0)} \big)R_{\beta\uparrow}
+ R_{\alpha\downarrow}^{(+)} R_{\beta\downarrow}
- R_{\alpha\uparrow}R_{\beta\uparrow}^{(-)}
+ R_{\alpha\downarrow}\big(R_{\beta\downarrow}^{(-)}
 -\sqrt{2}\,R_{\beta\uparrow}^{(0)}\big) \big] \\
{-}2:\!& \tfrac{\mathrm{i}}{\sqrt{2}} \mathrm{e}^{-2\mathrm{i}\varphi}
\big( {-}R_{\alpha\downarrow}^{(+)} R_{\beta\uparrow}
- R_{\alpha\downarrow} R_{\beta\uparrow}^{(-)} \big)
\end{array} \right. \\
\hat{\vec{L}}\psi_\beta &= -\mathrm{i}(\vec{r}\times\vec{\nabla})\psi_\beta
= \Bigg\{\!\! \begin{array}{rl}
{+}1:\!& \mathrm{e}^{\mathrm{i}\varphi}\big( \tfrac{\varrho}{\sqrt{2}}R_\beta^{(0)}
 + z R_\beta^{(+)} \big) \\
0:\!& \tfrac{\varrho}{\sqrt{2}}\big(R_\beta^{(+)}+R_\beta^{(-)}\big) \\
{-}1:\!& \mathrm{e}^{-\mathrm{i}\varphi}\big( \tfrac{\varrho}{\sqrt{2}}R_\beta^{(0)}
 - z R_\beta^{(-)} \big) \end{array}
\end{align}
\end{small}
\end{subequations}

In the actual calculation, it is necessary to choose projection of angular momentum $\mu$ and parity $\pi$ (together denoted also as $K^\pi$, and $\mu=\pm K$). Transition operators have the form of
\begin{equation}
\hat{M}_{\lambda\mu} = \sum_i M_{\lambda\mu}(\varrho_i,z_i)\,
\mathrm{e}^{\mathrm{i}\mu\varphi_i}
\end{equation}
where $M_{\lambda\mu}(\varrho,z)$ contains a function (or even derivatives) not dependent on $\varphi$. Choice of the two-quasiparticle pairs is restricted by $m_\alpha-m_\beta=\mu$ (time-reversed states are omitted from the discussion here for simplicity; in short, it is necessary to add pairs $(\alpha,\bar{\beta})$ with $m_\alpha+m_\beta=\mu$ during the duplicate removal similar to (\ref{order2qp})). Commutators are then evaluated in quasiparticle vacuum as
\begin{equation}
\langle[\hat{A}^\dagger,\hat{B}]\rangle = \frac{\gamma_T^A-\gamma_T^B}{2}
\sum_{\alpha\beta}u_{\beta\alpha}^{(\gamma_T^A)} u_{\alpha\beta}^{(\gamma_T^B)}
\langle\beta|\hat{A}^\dagger|\alpha\rangle\langle\alpha|\hat{B}|\beta\rangle
= \frac{1-\gamma_T^A\gamma_T^B}{2}\sum_{\alpha\beta}
u_{\alpha\beta}^{(\gamma_T^A)} u_{\alpha\beta}^{(\gamma_T^B)}
\langle\alpha|\hat{A}|\beta\rangle^*\,\langle\alpha|\hat{B}|\beta\rangle
\end{equation}

Single-particle operators (including densities and currents) can be expressed in terms of quasiparticles
\begin{align}
\hat{A} &= \frac{1}{2}\sum_{\alpha\beta}u_{\alpha\beta}^{(\gamma_T^A)}
\langle\alpha|\hat{A}|\beta\rangle
\big(\hat{\alpha}_\alpha^+\hat{\alpha}_{\bar{\beta}}^+ + \gamma_T^A
\hat{\alpha}_{\bar{\alpha}}^{\phantom{*}}\hat{\alpha}_{\beta}^{\phantom{|}}\big) \\
\label{axial_dens}
\hat{\mathbf{J}}_d(\vec{r}) &= \frac{1}{2}\sum_\mu\sum_{\alpha\beta\in\mu}
\mathbf{J}_{d;\alpha\beta}(\varrho,z)
\big(\hat{\alpha}_\alpha^+\hat{\alpha}_{\bar{\beta}}^+ + \gamma_T^d
\hat{\alpha}_{\bar{\alpha}}^{\phantom{*}}\hat{\alpha}_{\beta}^{\phantom{|}}\big)
\mathrm{e}^{-\mathrm{i}\mu\varphi},
\qquad\textrm{where }\ m_\alpha-m_\beta=\mu
\end{align}
Expression (\ref{axial_dens}) is defining the shorthand notation $\mathbf{J}_{d;\alpha\beta}(\varrho,z)$ for matrix elements of densities and currents which can have scalar, vector or tensor character. RPA phonons are defined as
\begin{equation}
\label{RPA_ax}
\hat{C}_\nu^+ = \frac{1}{2}\sum_{\alpha\beta}\big(
c_{\alpha\beta}^{(\nu-)} \hat{\alpha}_\alpha^+ \hat{\alpha}_{\bar{\beta}}^+
- c_{\alpha\beta}^{(\nu+)} \hat{\alpha}_{\bar{\alpha}} \hat{\alpha}_\beta \big)
\end{equation}
(factor $1/2$ is due to double counting of $\alpha\beta$ vs.~$\bar{\beta}\bar{\alpha}$) and their commutator with hermitian density/current operator is
\begin{equation}
\label{comm2_ax}
\langle[\hat{\mathbf{J}}_d(\vec{r}),\hat{C}_\nu^+]\rangle = \frac{1}{2}\sum_{\alpha\beta}
u_{\alpha\beta}^{(\gamma_T^d)} \langle\alpha|\hat{\mathbf{J}}_d(\vec{r})|\beta\rangle^*\,
\big( c_{\alpha\beta}^{(\nu-)} + \gamma_T^d c_{\alpha\beta}^{(\nu+)} \big)
= \frac{1}{2}\sum_{\alpha\beta}\mathbf{J}_{d;\alpha\beta}^\dagger(\varrho,z)
\big( c_{\alpha\beta}^{(\nu-)} + \gamma_T^d c_{\alpha\beta}^{(\nu+)} \big)\mathrm{e}^{\mathrm{i}\mu\varphi}
\end{equation}
where hermitian conjugation is understood in the sense of (\ref{hermit}) for vector/tensor components (see also decomposition of matrix elements (\ref{axial_me}) to these components) and the factor $\mathrm{e}^{\mathrm{i}\mu\varphi}$ will get canceled by $\mathrm{e}^{-\mathrm{i}\mu\varphi}$ from another $\hat{\mathbf{J}}_{d'}(\vec{r})$ in Skyrme interaction or in Coulomb integral (\ref{coul_ax}). The matrices $A$ and $B$ coming from the RPA equations $[\hat{H},\hat{C}_\nu^+] = E_\nu\hat{C}_\nu^+$ (\ref{fullRPA_eq}) are then
\begin{subequations}
\begin{align}
\label{axial_fullRPA_A}
A_{pp'} & = \delta_{pp'}\varepsilon_p + \sum_{dd'}
\iint\mathrm{d}\vec{r}_1\mathrm{d}\vec{r}_2
\frac{\delta^2\mathcal{H}}{\delta J_d(\vec{r}_1)\delta J_{d'}(\vec{r}_2)}
\mathbf{J}_{d;p}^\dagger(\varrho_1,z_1)\cdot\mathbf{J}_{d';p'}(\varrho_2,z_2) \\
\label{axial_fullRPA_B}
B_{pp'} & = \sum_{dd'} \gamma_T^d
\iint\mathrm{d}\vec{r}_1\mathrm{d}\vec{r}_2
\frac{\delta^2\mathcal{H}}{\delta J_d(\vec{r}_1)\delta J_{d'}(\vec{r}_2)}
\mathbf{J}_{d;p}^\dagger(\varrho_1,z_1)\cdot\mathbf{J}_{d';p'}(\varrho_2,z_2)
\end{align}
\end{subequations}
Index $p$ labels the $2qp$ pair (e.g.~$\alpha\beta$), satisfying $m_\alpha-m_\beta=\mu$ and the scalar product is understood in the spherical-tensor sense
\begin{equation}
\mathbf{A}^\dagger\cdot\mathbf{B} = \sum_s (-1)^s \big[\mathbf{A}^\dagger\big]_{-s}\big[\mathbf{B}\big]_s = \sum_s \big[\mathbf{A}\big]_s^*\,\big[\mathbf{B}\big]_s
\end{equation}

Pairing is included only for $\mu=0$ and positive parity, and only for diagonal $2qp$ pairs ($\beta=\alpha$). Density dependent pairing force then leads to two terms in the residual interaction: $\delta^2\mathcal{H}/\delta\kappa\delta\kappa$ and $\delta^2\mathcal{H}/\delta\kappa\delta\rho$.

Finally, a few words have to be said about numerical integration, in particular about Coulomb integral. Calculations are done on an equidistant cylindrical grid (with the grid spacing denoted here as $\Delta$). In that case, the radial integral has a non-zero (first) Euler-Maclaurin correction in the center (in contrast with spherical integration, which has none):
\begin{equation}
\label{axial_int}
\int_0^\infty f(\varrho)\varrho\,\mathrm{d}\varrho = \Big[
\tfrac{\Delta}{12}f(0) + 1\Delta f(1\Delta) + 2\Delta f(2\Delta) + 3\Delta f(3\Delta)
+\ldots \Big]\Delta
\end{equation}
Coulomb integral is
\begin{align}
\iint \frac{\rho_1^*(\varrho_1,z_1)\rho_2(\varrho_2,z_2)}{|\vec{r}_1-\vec{r}_2|}
&\,\mathrm{e}^{\mathrm{i}m_1\varphi_1-\mathrm{i}m_2\varphi_2}\mathrm{d}\vec{r}_1
\mathrm{d}\vec{r}_2 = 2\pi\delta_{m_1m_2}
\int_0^\infty\rho_1^*(\varrho_1,z_1)\varrho_1\mathrm{d}\varrho_1\mathrm{d}z_1
\int_0^\infty\rho_2(\varrho_2,z_2)\varrho_2\mathrm{d}\varrho_2\mathrm{d}z_2 \nonumber\\
\label{coul_ax}
&\qquad\qquad\qquad\qquad\qquad
 \times \frac{g_m\Big(\tfrac{2\varrho_1\varrho_2}{(z_1-z_2)^2
+\varrho_1^2+\varrho_2^2}\Big)}{\sqrt{(z_1-z_2)^2+\varrho_1^2+\varrho_2^2}} \\
\label{coul_ax_g}
\textrm{where }\ g_m(x) &= \int_{-\pi}^{\pi}
\frac{\cos(m\varphi)}{\sqrt{1-x\cos\varphi}} \mathrm{d}\varphi
= 2\pi\sum_{k=0}^{\infty} \frac{(4k+2m-1)!!}{k!(k+m)!}\bigg(\frac{x}{4}\bigg)^{m+2k}
\end{align}
Function $g_m(x)$ is evaluated by numerical integration or by Taylor expansion (for small $x$ and large $m$). It has a logarithmic singularity in $x\to1^{-}$ like $g_m(1-t) = (O(t)-\sqrt{2})\ln t + O(1)$. The singularity can be replaced in the equidistant cylindrical grid by an empirical expression (found by investigation of the convergence with decreasing grid spacing):
\begin{equation}
\label{coul_ax_0}
\frac{g_m\Big(\tfrac{2\varrho_1\varrho_2}{(z_1-z_2)^2+\varrho_1^2+\varrho_2^2}\Big)}
{\sqrt{(z_1-z_2)^2+\varrho_1^2+\varrho_2^2}} \bigg|_{z_1=z_2,\,\varrho_1=\varrho_2}
= \frac{1}{\varrho} \bigg[2\ln\frac{\varrho}{\Delta} + 6.779948935
- 4\sum_{n=1}^m \frac{1}{2n-1} \bigg] + O(\Delta^2)
\end{equation}
For the point on the axis ($z_1=z_2$ and $\varrho_1=\varrho_2=0$, assuming $m=0$; otherwise the contribution is zero), the first term in (\ref{axial_int}) is replaced as
\begin{equation}
\frac{\Delta}{12} \frac{\rho_2(0,z_2) g_0(0)}{\sqrt{0^2+0^2+0^2}}
\quad\mapsto\quad 2.1770180559\,\rho_2(0,z_1)
\end{equation}

\section{Numerical results}
\begin{figure}[h]
\begin{center}
\includegraphics[width=1.0\textwidth]{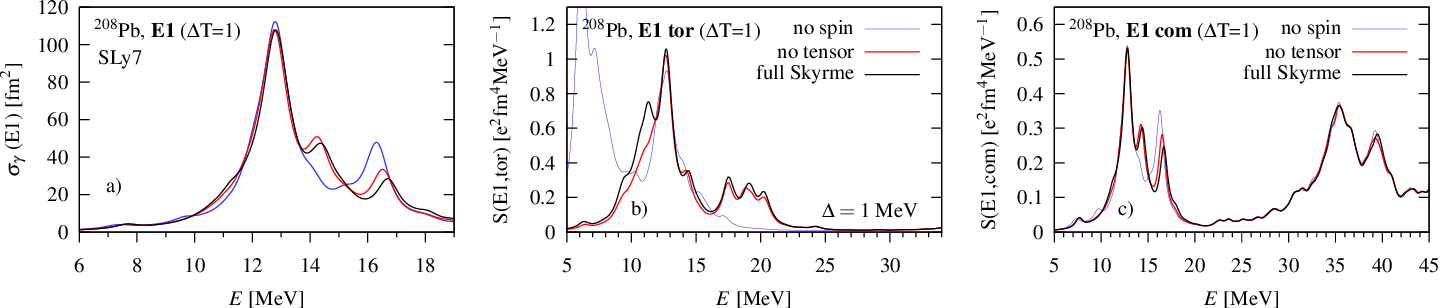}
\end{center}
\vspace{-15pt}
\caption{Isovector E1 resonances in $^{208}$Pb for parametrization SLy7 \cite{Chabanat1998} with corresponding effective charges: a) giant dipole (long-wave), $z_p=N/A,\,z_n=-Z/A$; b) toroidal and c) compression resonances with $z_p=-z_n=0.5,\,g_p=-g_n=4.7\varsigma$ ($\varsigma=0.7$).}\label{fig_E1T1}
\end{figure}

\begin{figure}[h]
\begin{center}
\includegraphics[width=0.7\textwidth]{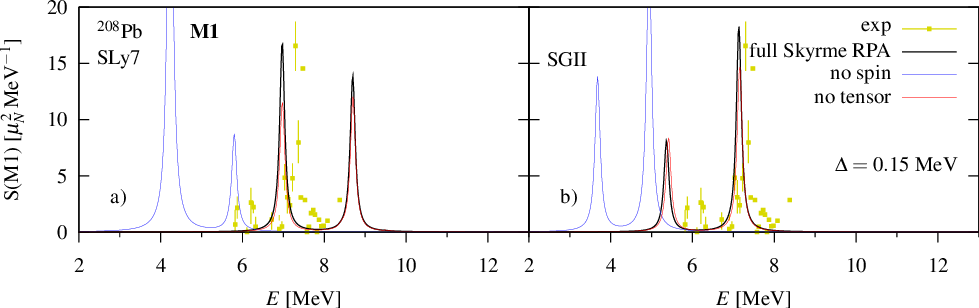}
\end{center}
\vspace{-15pt}
\caption{M1 resonances in $^{208}$Pb with natural charges and parametrizations: a) SLy7 \cite{Chabanat1998}, b) SGII \cite{Giai1981}. Experimental data are from \cite{Laszewski1988}.}\label{fig_M1}
\end{figure}

The influence of omission of tensor and spin terms in Skyrme functional (\ref{Skyrme_DFT}) is illustrated for spherical nucleus $^{208}$Pb. Skyrme parametrization SLy7 \cite{Chabanat1998}, fitted with tensor term, was utilized. Isovector E1 strength functions (long-wave, toroidal and compression) are given in Fig.~\ref{fig_E1T1} with Lorentzian smoothing of 1 MeV. Magnetic M1 strength functions are given in Fig.~\ref{fig_M1} (smoothing 0.15 MeV).

As can be seen, omission of tensor terms has only a minor influence on all results, while the omitted spin terms give a large unphysical shift mainly in M1 strength functions, and also in toroidal E1. Both E1 and M1 were calculated with exhaustive basis: 120 main shells of spherical harmonic oscillator (with length 1.7 fm) in HF, and 106 main shells passed to RPA. Spurious state in E1 is under 1 keV for the full Skyrme and ``no spin'' case, while the ``no tensor'' case gave imaginary state at 0.78i MeV due to broken self-consistency in residual interaction (tensor term was omitted only in RPA).

\begin{figure}[h]
\begin{center}
\includegraphics[width=0.8\textwidth]{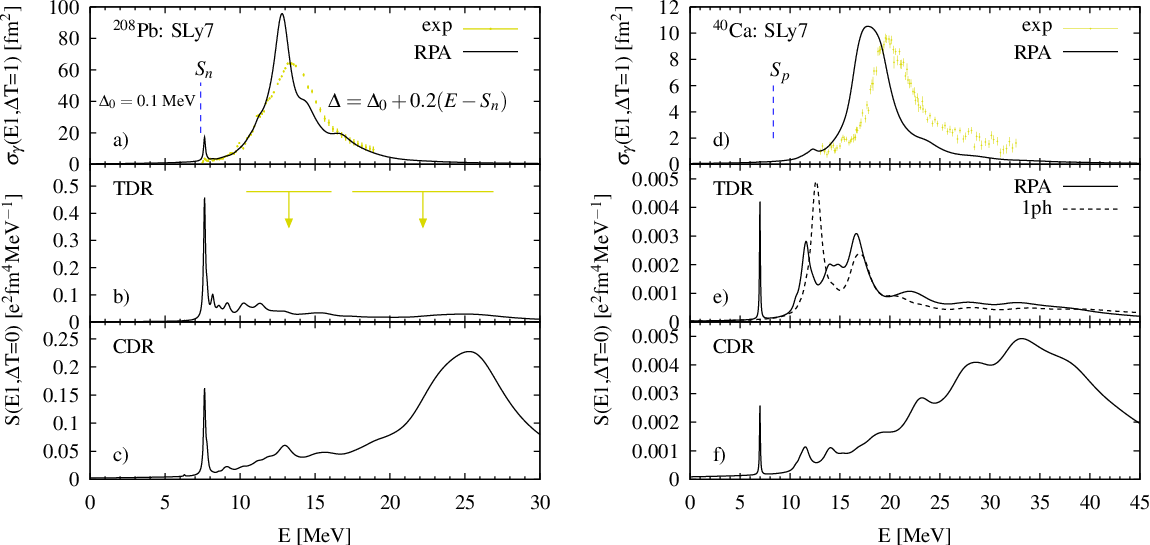}
\end{center}
\vspace{-15pt}
\caption{E1 strengths in $^{208}$Pb and $^{40}$Ca with pygmy region emphasized by variable smoothing. Dipolar resonances are compared with experiment a) \cite{Veyssiere1970}, b,c) \cite{Youngblood2004} (isoscalar E1), d) \cite{Ahrens1975}. Isoscalar toroidal (b,e) and compression (c,f) strength functions are calculated with $z_p=z_n=0.5,\,g_p=g_n=0.88\varsigma$ ($\varsigma=0.7$).}\label{fig_pygmy-sf}
\end{figure}
\begin{figure}[h]
\vspace{-5pt}
\begin{center}
\includegraphics[width=1.0\textwidth]{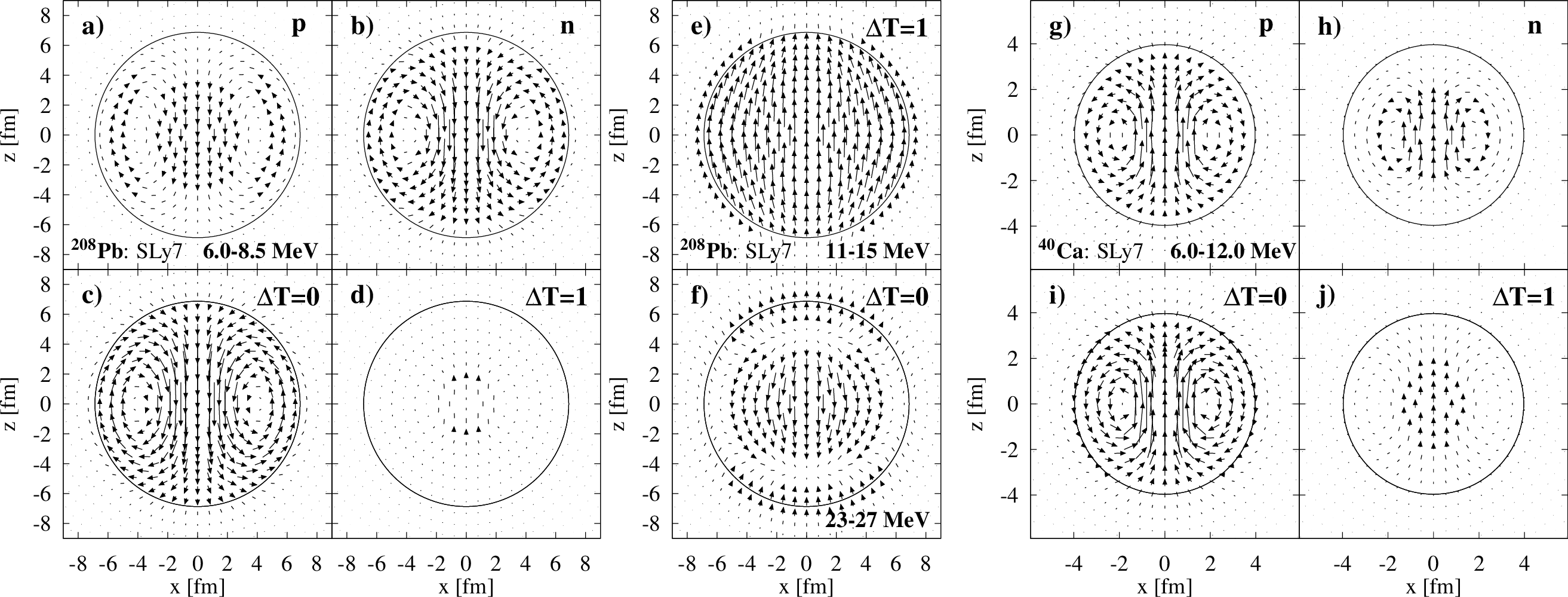}
\end{center}
\vspace{-15pt}
\caption{E1 transition currents in $^{208}$Pb and $^{40}$Ca in various energy intervals, weighted by isovector E1 operator. $\Delta$T=0 is the sum of proton and neutron current, $\Delta$T=1 gives their difference.}\label{fig_cur}
\end{figure}

As was shown in our previous work \cite{Repko2013,Reinhard2014}, pygmy resonance (low-lying part of the E1 resonance) has a pronounced isoscalar toroidal nature, contrary to the previous interpretations, which assumed vibration of neutron skin against the core of the nucleus. These conclusions are illustrated here for spherical $^{208}$Pb and $^{40}$Ca by the strength function of isovector E1 (long-wave) and isoscalar toroidal and compression E1, with variable smoothing: starting with $\Delta=0.1\ \mathrm{MeV}$ up to particle emission threshold, and then linearly increasing to approximately simulate the escape width and coupling to complex configurations. Fig.~\ref{fig_pygmy-sf} shows that pygmy resonance is not visible in isovector E1 for symmetric nucleus $^{40}$Ca in agreement with its isoscalar nature (as found also in \cite{Papakonstantinou2011}). Strength functions and Fig.~\ref{fig_cur} demonstrate that the toroidal flow is dominant in low-energy region, while the compression flow (more akin to surface vibration) dominates for higher energies.

\begin{figure}[h]
\begin{center}
\includegraphics[width=0.7\textwidth]{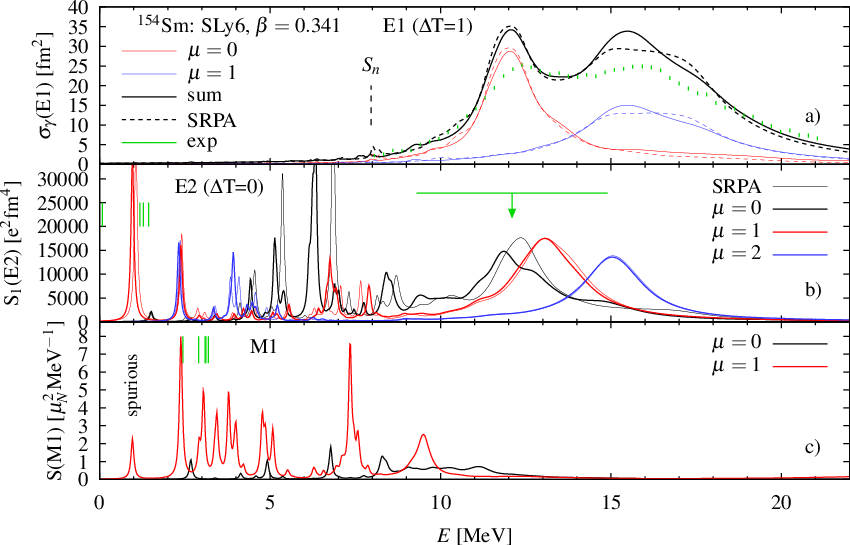}
\end{center}
\vspace{-15pt}
\caption{Giant resonances in deformed nucleus $^{154}$Sm given with variable smoothing, and showing pronounced deformation splitting as a function of $\mu$. a) E1 resonance with SRPA denoted by dashed lines + experiment \cite{Carlos1974}, b) E2 resonance with SRPA denoted by thin lines + experiment \cite{Youngblood2004}, c) M1 resonance, showing spurious peak at 0.96 MeV, corresponding to collective rotation.}\label{fig_Sm154}
\end{figure}
\begin{figure}[t]
\begin{center}
\includegraphics[width=0.5\textwidth]{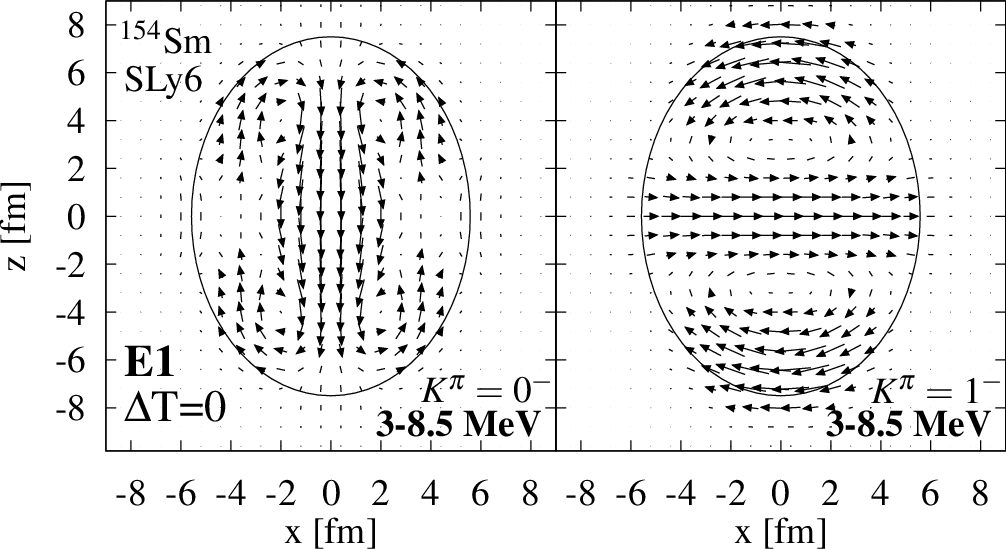}
\end{center}
\vspace{-15pt}
\caption{Isoscalar E1 transition currents in the pygmy region of $^{154}$Sm by full RPA.}\label{fig_Sm-pygmy}
\end{figure}

Finally, a calculation with deformed nucleus $^{154}$Sm is shown to demonstrate the accuracy of SRPA vs.~full RPA (see Fig.~\ref{fig_Sm154}), with a grid spacing of 0.4 fm, using single-particle states up to 40 MeV in RPA. The slowest calculations were for $\mu=1$ (E1 and E2) with around 22.6 thousand $2qp$ pairs, taking around 24 hours utilizing 8 threads and around 23 GB of RAM for single $\lambda\mu$. Pairing strengths $V_q$ for SLy6 were taken from \cite{Guo2007}. Slight differences between full and separable RPA can be explained by an imperfect coverage of the residual interaction by its separable approximation (generated by five operators in each case). Toroidal currents are present in the pygmy region of $^{154}$Sm E1 resonance, as is shown in Fig.~\ref{fig_Sm-pygmy}. 

\section{CONCLUSIONS}
Skyrme RPA formalism and some numerical results are presented for the case of spherical and axial symmetry. Formulae in axial basis are given here in detail as a reference for further calculations with full RPA, which is now becoming feasible on the common workstations.

Importance of time-odd Skyrme terms was shown for magnetic and toroidal transitions. Toroidal nature of pygmy resonance was demonstrated by transition current maps for both spherical ($^{40}$Ca, $^{208}$Pb) and deformed nuclei ($^{154}$Sm).

\noindent \\ACKNOWLEDGMENT:\\ 
This work was supported by the Czech Science Foundation (P203-13-07117S) and by Votruba-Blokhintsev grant (Czech Republic-BLTP JINR).

\end{document}